\documentclass{article}
\usepackage{amsmath}
\usepackage{amssymb}
\usepackage[margin=2cm]{geometry}
\usepackage{graphicx}
\usepackage{cite}
\usepackage{color}

\newcommand{\qexpect}[1]{\langle #1 \rangle_{q}}%
\newcommand{\trace}[1]{\mathrm{Tr}\left( #1\right)} %
\newcommand{\norder}[1]{:\hspace{-3.0pt}#1\hspace{-3.0pt}:}% normal ordered quantity
\newcommand{\nH}[1]{:\hspace{-3.0pt}H(#1)\hspace{-3.0pt}:}% normal ordered Hamiltonian :H():
\newcommand{\nHM}{:\hspace{-3.0pt}H^M\hspace{-3.0pt}:}% normal ordered Hamiltonian :H^M:
\newcommand{\Veff}{V_{\mathrm{eff}}}% normal ordered Hamiltonian :H^M:
\newcommand{\Tph}{T_{\mathrm{ph}}}%physical temperature

\newcommand{\figsize}{0.45\textwidth}

\title{Phase transition for the system of small volume in the $\phi^4$ theory in the Tsallis nonextensive statistics}
\author{Masamichi Ishihara}
\begin{document}
\maketitle
%%%%%%%%%%%%%%%%%%%%%%%
\begin{abstract}
We studied the effects of the nonextensivity on the phase transition for the system of small volume $V$ in the $\phi^4$ theory 
in the Tsallis nonextensive statistics of entropic parameter $q$ and temperature $T$, 
when the deviation from the Boltzmann-Gibbs statistics, $|q-1|$, is small.
We calculated the condensate and the mass to the order $q-1$ 
with the normalized $q$-expectation value under the massless free particle approximation.
%%% 
%%% 
The following facts were found. 
The condensate $\Phi$ divided by $v$,  $\Phi/v$, at $q$ is smaller than that at $q'$ for $q>q'$ as a function of $\Tph/v$ 
which is the physical temperature $\Tph$ divided by $v$,  
where $\Tph$ at $q=1$ coincides with $T$ and $v$ is the value of the condensate at $T=0$.
The mass decreases, reaches minimum, and increases after that, as $\Tph$ increases. 
The mass at $q>1$  is lighter than the mass at $q=1$ at low physical temperature and heavier than the mass at $q=1$ at high physical temperature. 
The effects of the nonentensivity on the physical quantity as a function of $\Tph$ become strong as $|q-1|$ increases. 
%%%%%
The results indicate 
the significance of the definition of the expectation value, the definition of the physical temperature, and the constraints for the density operator, 
when the terms including the volume of the system are not negligible.
%%% 
\end{abstract}

%%%%%%%%%%%%%%%%%%%%%%%
\section{Introduction}
%%% べき的な分布があることと、物理量が計算されていること
A power-like distribution appears and is of interest in various branches of science. 
A momentum distributions in a high energy collision shows a power-like distribution, 
and the distribution is described well by a Tsallis distribution which has an entropic parameter $q$ 
\cite{Alberico2009, Cleymans2012, Marques2015, GS2015, Zheng2016, Lao2016,Cleymans2017-WoC, Cleymans2017,  Osada-Ishihara-2017, Bhattacharyya2017-prepri}. 
Therefore, scientists have calculated the physical quantities under a Tsallis distribution, 
such as themodynamic quantities \cite{Lavagno2002, Bhattacharyya2016}, %%
fluctuation and  correlation \cite{Osada-Ishihara-2017, Alberico2000, Ishihara2017, Ishihara2017-inpress}, etc. 

%%% ツァリス統計が導入されていること
The statistical mechanics called 'Tsallis nonextensive statistics' was proposed
to describe the phenomena which show power-like distributions, 
and has been applied to the various phenomena \cite{Book:Tsallis}.
The nonextensivity is measured by the quantity $q-1$, and 
the effects of the nonextensivity have been studied. 
The definition of the expectation value in the Tsallis nonextensive statistics 
differs from that in the Boltzmann-Gibbs (BG) statistics \cite{Tsallis1998,Aragao-PhysicaA2003}.
The third choice of the expectation value given in ref.~\cite{Tsallis1998} 
(the normalized $q$-expectation value \cite{Tsallis1998, Kalyana2000, Aragao-PhysicaA2003, Eicke-prepri}) 
is physically preferable, 
and the Tsallis nonextensive statistics with the expectation value has been applied to many phenomena.

%%% 相転移 at high energies
The Tsallis nonextensive statistics has been applied to the phenomena at high energies, and 
an interesting topic at high energies is the phase transition. 
Chiral phase transition was studied with the Nambu-Jona-Lasinio model \cite{Rozynek2009, Rozynek2016}
and the linear sigma model \cite{Ishihara2015, Ishihara2016, Shen2017}.
The study of the phase transition is a significant topic when the momentum distribution is described well by a Tsallis distribution.

%%% 有限体積のこと
An application in a field theory is the calculation of the propagator 
within the framework of the Tsallis nonextensive statistics \cite{Kohyama-Tsallis06}.
The researchers dealt with the system of finite volume in the Tsallis nonextensive statistics of quite small $|q-1|$ in the study. 
The terms including the volume may affect the quantities, 
and it may be worth to estimate the effects of the terms for the system of small volume.

%%% 目的
The purpose of the present paper is to study the effects of the nonextensivity on the phase transition in the $\phi^4$ theory. 
We adopt the normalized $q$-expectation value with the density operator in the Tsallis nonextensive statistics of small $|q-1|$ for the system of small volume.
The condensate and the mass are calculated as a function of the temperature for various $q$, 
and the critical temperature is estimated.

%%% 結果
We summarize the results briefly.
The condensate decreases and reaches zero as the temperature increases.
The condensate $\Phi$ divided by $v$,  $\Phi/v$, at $q$ is smaller than that at $q'$ for $q>q'$ 
as a function of $\Tph/v$ which is the physical temperature $\Tph$ divided by $v$, 
where $v$ is the value of the condensate at $T=0$.
The mass at $q>1$  is lighter than the mass at $q=1$ at low physical temperature,  and heavier than the mass at $q=1$ at high physical temperature. 
These results indicate that 
the definition of the expectation value, the definition of the physical temperature, and the constraints for the density operator are significant.

%%%%%%%%%%%%%%%
This paper is organized as follows.
%%%%%
In section \ref{sec:theory}, 
we employ the $\phi^4$ theory in the Tsallis nonextensive statistics.
The critical temperature, the condensate, and the mass are calculated in the Tsallis nonextensive statistics 
with the normalized $q$-expectation value for the system of small volume.  
%%%%%
In section \ref{sec:numericalcalc}, 
the condensate and the mass are numerically estimated.
The temperature dependences are shown for various $q$ without or with the term including the volume of the system.
The critical temperature can be estimated from these calculations.  
The last section are assigned for discussion and conclusion.
%%%%%

%%%%%%%%%%%%%%%%%%%%%%%
\section{Nonextensive effects in the $\phi^4$ theory}
\label{sec:theory}

%%\subsection{The expectation value in the Tsallis nonextensive statistics}
\subsection{Brief review of the Tsallis nonextensive statistics}

In this subsection, we review the Tsallis nonextensive statistics with the normalized $q$-expectation value briefly.
%%%%%
The density operator $\rho$ in the Tsallis nonextensive statistics is defined by 
\begin{align}
\rho := \frac{\rho_u}{\trace{\rho_u}} , \qquad \rho_u := \left[ 1-(1-q) \frac{\beta}{c_q} (H - \qexpect{H}) \right]^{1/(1-q)} , 
\end{align}
where $H$ is the Hamiltonian, $\beta$ is the inverse temperature,  
$q$ is the entropic parameter, $c_q$ is the $q$-dependent constant, 
and $\qexpect{H}$ is the expectation value of the Hamiltonian in the statistics. 
The definition of the expectation value (the normalized $q$-expectation value)  is different from that in the BG statistics. 
The normalized $q$-expectation value is defined by
\begin{align}
\qexpect{O} := \frac{\trace{\rho_u^q O}}{\trace{\rho_u^q}} .
\end{align}
We adopt the normalized $q$-expectation value in the present study because of physical relevance, $\qexpect{1}=1$. 

%%%
The following self-consistent equation should be satisfied from the definition of the expectation value:
\begin{align}
\qexpect{H} = \frac{\trace{\rho_u^q H}}{\trace{\rho_u^q}} , 
\label{eqn:self-consistent}
\end{align}
where $\qexpect{H}$ is included in the right-hand side of eq.~\eqref{eqn:self-consistent}.
The constant $c_q$ should also satisfy the following relation:   
\begin{equation}
c_q = \left( Z_q \right)^{1-q} , 
\label{eqn:cqZq}
\end{equation}
where $Z_q$ is the partition function defined by
\begin{equation}
Z_q = \trace{\rho_u} .
\end{equation}
The above equations are used in the following calculations.

%%%%%%%%%
\subsection{Application of the Tsallis nonextensive statistics to the $\phi^4$ theory}
We start with the Hamiltonian of the $\phi^4$ theory to calculate the effective potential at finite temperature
in the Tsallis nonextensive statistics. 
The Hamiltonian density is
\begin{align}
{\cal H}(\phi) = \frac{1}{2} (\partial^0 \phi)^2 + \frac{1}{2} (\nabla \phi)^2 + \frac{\lambda}{4} (\phi^2 - v^2)^2  . 
\end{align}
We shift the field $\phi$ as $\Phi + \varphi$ with $\langle 0 |  \phi | 0 \rangle = \Phi$,  where $|0\rangle$ is the vacuum state.
The Hamiltonian is given by 
\begin{subequations}
\begin{align}
{\cal H}(\phi) &= {\cal H}(\Phi) + {\cal H}_{\mathrm{linear}}(\varphi)   + {\cal H}_{\mathrm{int}}(\varphi)   + {\cal H}_{\mathrm{indep.}}(\varphi)   ,\\
& {\cal H}(\Phi)  = \frac{1}{2} (\partial^0 \Phi)^2 + \frac{1}{2} (\nabla \Phi)^2 + \frac{\lambda}{4} (\Phi^2 - v^2)^2  ,\\ 
& {\cal H}_{\mathrm{linear}}(\varphi)   = (\partial^0 \Phi)(\partial^0 \varphi) + (\nabla \Phi)(\nabla \varphi) + \lambda (\Phi^2 - v^2) \Phi \varphi ,\\
& {\cal H}_{\mathrm{int}}(\varphi) = \lambda \left\{ \frac{1}{2} \left(\Phi^2 - v^2\right) \varphi^2 + \Phi^2 \varphi^2 + \Phi \varphi^3 \right\} ,\\
& {\cal H}_{\mathrm{indep.}}(\varphi)  = \frac{1}{2} (\partial^0 \varphi)^2  + \frac{1}{2} (\nabla \varphi)^2 + \frac{\lambda}{4} \varphi^4  . 
\end{align}
\end{subequations}

%%%%%%%%
Hereafter, we use the normal ordered Hamiltonian with respect to the creation and annihilation operators of $\varphi$. 
The expectation value of the normal ordered Hamiltonian under the massless free particle approximation \cite{Gavin1994, Ishihara1999,Ishihara2017}
is given by 
%%%%%%%%%%%%%%%%%%%%%%%%%%%%%%%%
\begin{subequations}
\begin{align}
\qexpect{\nH{\phi}} 
&= \frac{1}{2} (\partial^0 \Phi)^2  + \frac{1}{2} (\nabla \Phi)^2 + \Veff(\Phi) % V_{eff}(\Phi) 
 - \frac{9}{4} \lambda \left( I(T,q) \right)^2 + \frac{3}{2} \lambda v^2 I(T,q) 
    + \qexpect{:\hspace{-3pt}{\cal H}_{\mathrm{indep.}}(\varphi) \hspace{-3pt}:}
,\\
& \Veff(\Phi) = \frac{\lambda}{4} \left[ \Phi^2 + 3 I(T,q) -v^2 \right]^2 
%% V_{eff}(\Phi) = \frac{\lambda}{4} \left[ \Phi^2 + 3 I(T,q) -v^2 \right]^2 
, 
\end{align}
\end{subequations}
where $T$ is the temperature, $T=\beta^{-1}$, and $I(T,q)$ is defined by 
\begin{align}
I(T,q) := \qexpect{\norder{\varphi^2}} . 
\end{align}
%%%%%%%%%%%%%%%%%%%%%%%%%%%%%%%%
%%%\begin{align}
%%%\qexpect{\nH{\phi}} 
%%%= {\cal H}(\Phi) + \frac{3}{2} \lambda I(T,q) \Phi^2 - \frac{1}{2} \lambda v^2 I(T,q) 
%%%    + \qexpect{:\hspace{-3pt}{\cal H}_{indep.}(\varphi) \hspace{-3pt}:},
%%%\end{align}
%%%%%%%%%%%%%%%%%%%%%%%%%
%%%The expectation value is rewrtten in the following form:

The critical temperature and the mass are determined from $\Veff$.
Therefore, we attempt to estimate $I(T,q)$ under the massless free particle approximation in the Tsallis nonextensive statistics.
We note that the last term $\qexpect{:\hspace{-3pt}{\cal H}_{\mathrm{indep.}}(\varphi) \hspace{-3pt}:}$ is independent of $\Phi$ 
in the present approximation.
The normal ordered Hamiltonian for a free scalar field is give by
%%%%%%%%
\begin{align}
\nHM = \sum_{\vec{l}}  \omega_{\vec{l}}  a^{\dag}_{\vec{l}}  a_{\vec{l}}  ,
\end{align}
%%%%%%%%
where $\omega_{\vec{l}}$ is the energy of a particle with momentum $\vec{l}$ 
and $a_{\vec{l}}$ is the annihilation operator.
We use the operator $\rho_u$ with the Hamiltonian ($\nHM$): 
\begin{align}
\rho_u^M = \left[  1-(1-q) \left( \frac{\beta}{c_q^M} \right) (\nHM - \qexpect{\nHM}) \right]^{1/(1-q)} ,
\end{align}
where we attach the superscript $'M'$ to $\rho_u$ and $c_q$ in order to clarify that the Hamiltonian $\nHM$ is used.
The self-consistent equation is rewritten as follows
\begin{align}
\qexpect{\nHM} = 
\frac{\trace{\left[  1-(1-q) \left( \frac{\beta}{c_q^M} \right) (\nHM - \qexpect{\nHM}) \right]^{q/(1-q)}   \nHM}}
{\trace{\left[  1-(1-q) \left( \frac{\beta}{c_q^M} \right) (\nHM - \qexpect{\nHM}) \right]^{q/(1-q)}}}
.
\label{eqn:self-conf:M}
\end{align}
%%%%%%%%

In this study, we focus on the system of small $|q-1|$. 
For simplicity, we use the variable $\varepsilon = 1-q$.
We expand $\qexpect{\nHM}$ and $c_q^M$ as follows:
%%%%%%%%
\begin{subequations}
\begin{align}
& \qexpect{\nHM} = E_0^{M} - \varepsilon E_1^M + O(\varepsilon^2) \label{expand:E} ,\\
& c_q^M = c_0^{M} - \varepsilon c_1^M + O(\varepsilon^2) \label{expand:c} ,\\
& Z_q^M = Z_0^M - \varepsilon Z_1^M + O(\varepsilon^2)  \label{expand:Z} .
\end{align}
\end{subequations}
%%%%%%%%
We have the following relations from eq.~\eqref{eqn:cqZq}:
\begin{subequations}
\begin{align}
c_0^M &= 1,\\
c_1^M & = - \ln Z_0^M = -\ln \trace{\exp[-\beta(\nHM-E_0^M)]}  .
\end{align}
\end{subequations}
%%%%%%%%%%%%%%%%%%%
The quantity $(\rho_u^M)^q$ is expanded as follows:
\begin{align}
(\rho_u^M)^q = e^{\beta E_0^M} e^{-\beta (:H^M:)} \left\{ 1 + \varepsilon \left[ L_0 + L_1 (\nHM) + L_2 (\nHM)^2 \right] + O(\varepsilon^2) \right\}
,
\end{align}
where $L_0$, $L_1$, and $L_2$ are defined by 
\begin{subequations}
\begin{align}
L_0 &= - \beta \left[ (1-c_1^M) E_0^M + E_1^M + \frac{1}{2} \beta (E_0^M)^2 \right] ,\\
L_1 &= \beta \left[ (1-c_1^M) + \beta E_0^M \right] ,\\
L_2 &= -\frac{1}{2} \beta^2  .
\end{align}
\end{subequations}

%%%%%%%%%%%%%%%%%%%
%%%%%%%% Inserting eqs.~\eqref{expand:E} and \eqref{expand:c} into eq.~\eqref{eqn:self-conf:M} and comparing the terms to the $O(\varepsilon)$,   
From eq.~\eqref{eqn:self-conf:M} , 
we obtain 
\begin{align}
& \beta \qexpect{\nHM} = \beta E_0^{M} - \varepsilon \beta E_1^M + O(\varepsilon^2) 
= \left\{ \frac{\pi^2}{30}  + (q-1) \left[ \frac{\pi^2}{5} - \frac{4 \pi^4}{675} (\beta^{-3} V) \right] \right\} (\beta^{-3} V), 
%%\\
%%& c_1^M = - \frac{2\pi^2}{45}  (\beta^{-3} V),  
\end{align}
where $V$ is the volume of the system. 
The second term in the square bracket comes from the coefficient $c_1^M$.

%%%% 
The quantity $I(T,q)$ is required to calculate $\Veff$.  %%%  $\equiv \qexpect{\norder{\varphi^2}}$ 
The quantity $I(T,q)$ to the $O(\varepsilon)$ is given by 
\begin{align}
%%% \qexpect{\norder{\varphi^2}}
I(T,q)
= & V^{-1} \sum_{\vec{k}} \left( \frac{1}{\omega_{\vec{k}} }\right)  \left( \frac{\Lambda_{0,\vec{k}}}{\Lambda_{0}}  \right) 
   + \varepsilon L_1 V^{-1} \sum_{\vec{k}} \left( \frac{1}{\omega_{\vec{k}} }\right) \left[ 
     \left( \frac{\Lambda_{1,\vec{k}}}{\Lambda_{0}}  \right) - \left( \frac{\Lambda_{1}}{\Lambda_{0}}  \right) \left( \frac{\Lambda_{0,\vec{k}}}{\Lambda_{0}}  \right) 
       \right]
   \nonumber \\ & \qquad
   + \varepsilon L_2 V^{-1} \sum_{\vec{k}} \left( \frac{1}{\omega_{\vec{k}} } \right) \left[ 
     \left( \frac{\Lambda_{2,\vec{k}}}{\Lambda_{0}}  \right) - \left( \frac{\Lambda_{2}}{\Lambda_{0}}  \right) \left( \frac{\Lambda_{0,\vec{k}}}{\Lambda_{0}}  \right) 
       \right]
   ,
\end{align}
where $\Lambda_n$ and $\Lambda_{n,\vec{k}}$ are defined by 
%%%%%%%%%%%%%%%
\begin{subequations}
\begin{align}
& \Lambda_{n} := \trace{ \exp(-\beta \nHM) (\nHM)^n } ,
\\
& \Lambda_{n,\vec{k}} := \trace{ \exp(-\beta \nHM ) (\nHM)^n a^{\dag}_{\vec{k}} a_{\vec{k}}} .
\end{align}
\label{Lambda:main}
\end{subequations}
%%%%%%%%%%%%%%%
These quantities, $\Lambda_n$ and $\Lambda_{n,\vec{k}}$, are given explicitly in the appendix~\ref{app:sec:trace}.
%%%%%%%%%%%%%%%
The quantity $I(T,q)$ is calculated     %%%$\equiv \qexpect{\norder{\varphi^2}}$ 
with the help of the results of the integrals given in the appendix~\ref{app:sec:integrals},  we have
\begin{subequations}
\begin{align}
%%% \equiv  \qexpect{\norder{\varphi^2}}  
I(T,q) & = \frac{1}{12\beta^2} + \frac{(q-1)}{\beta^2} \left[\frac{1}{12} - \frac{\pi^2}{135} \left( \beta^{-3} V\right) \right] 
\label{ITq:a}
\\%\nonumber 
&= \frac{q}{12\beta^2}  - \frac{\pi^2}{135} \frac{(q-1)}{\beta^5}  V
.
\label{ITq:b}
\end{align}
\end{subequations}
%%%%%%%%
The  quantity $I(T,q=1)$ is the well-known result.
We obtain a simple result when the last term of eq.~\eqref{ITq:b} is negligible: $I(T,q) = qT^2/12$.

%%%%%%%%%%%%%%%%%%
We attempt to estimate the critical temperature,  the condensate, and the mass. 
We now consider the case that the second term of eq.~\eqref{ITq:b} is small compared with the first term of  eq.~\eqref{ITq:b}. 
For example, 
the absolute value of the ratio of the second term of  eq.~\eqref{ITq:b} to the first term of eq.~\eqref{ITq:b} 
is less than 0.25 for  $v^3 V = 0.2$, $T/v=2.5$, and $q=1.1$. 
%%%%%%%%%%%
The critical temperature $T_c(q)$ is approximately  estimated as 
\begin{equation}
T_c(q) \sim v \left[   \frac{2}{\sqrt{q}}  + \frac{(q-1)}{q^3} \left( \frac{96\pi^2}{135} \right)  (v^3 V) \right] .
\label{eqn:Tc:V}
\end{equation}
%%%%%%%%%%%
The condensate $\Phi(T,q)$ is given by 
\begin{align}
\Phi(T,q) = \left\{ 
\begin{array}{ll}
\pm v \sqrt{1 - 3 I(T,q)/v^2} & \qquad (T < T_c) \\
0 & \qquad (T \ge T_c) 
\end{array}
\right.
.
\end{align}
%%%%%%%%%%%
The mass $m(T,q)$ is given by 
\begin{align}
m(T,q) = \left\{ 
\begin{array}{ll}
v \sqrt{2 \lambda \left( 1 - 3 I(T,q)/v^2\right) } & \qquad (T < T_c) \\
v \sqrt{\lambda \left( 3 I(T,q)/v^2 - 1\right) }  & \qquad (T \ge T_c) 
\end{array}
\right. 
. 
\end{align}
%%%%%%%%%%%%%%%%%%

%%%%%%%%%%%
The critical temperature, the condensate, and the mass are expressed as follows, 
when the term including $V$ in eq.~\eqref{ITq:b} is negligible.
The critical temperature $T_c(q)$ is simply
\begin{equation}
T_c(q) = \frac{2v}{\sqrt{q}} .
\label{eqn:Tc}
\end{equation}
%%%%%%%%
%%%%%%%%%%%%%%%%%%%%%%%%
The condensate is given by 
\begin{equation}
\left| \Phi(T,q) \right| = \left\{ 
\begin{array}{ll}
v \sqrt{1 - \frac{qT^2}{4v^2}} & \qquad (T<T_c)\\
0 & \qquad (T \ge T_c)
\end{array}
\right.
. 
\label{eqn:Phi}
\end{equation}
%%%%%%%%
%%%%%%%%%%%%%%%%%%%%%%%%%%%%%%%%%%%%
The mass $m$ is given by 
\begin{equation}
m(T,q) = \left\{ 
\begin{array}{ll}
v \sqrt{2\lambda \left( 1 - \frac{qT^2}{4v^2} \right) }  & \qquad (T<T_c)\\
v \sqrt{\lambda \left( \frac{qT^2}{4v^2}  - 1 \right)}   & \qquad (T \ge T_c)
\end{array}
\right.
.
\label{eqn:mass}
\end{equation}
%%%%%%%%%%%%%%%
The $q$-dependences of the quantities, $\Phi(T,q)$ in eq.~\eqref{eqn:Phi} and $m(T,q)$ in eq.~\eqref{eqn:mass}, 
are absorbed by the effective temperature $T^* :=\sqrt{q} T$.
%%% when the terms including $V$ are negligible.

%%%%%%%%%%%%%%%%%%%%%%%%%%%%%%%%%%%%%%%%%%
%%% 
The temperature called physical or effective temperature $T_{\mathrm{ph}}$ \cite{Aragao-PhysicaA2003, Eicke-prepri, Kalyana2000} is 
defined to analyze the effects of the nonextensivity:
\begin{equation}
\Tph := c^M_q T .
\end{equation}
The physical temperature $T_{\mathrm{ph}}$ is described as
\begin{equation}
\Tph
= \left( c_0^M - \varepsilon c_1^M \right) T  
= \left( 1 +(q-1) \left(- \frac{2\pi^2}{45} \right) T^3 V \right) T  
.
\label{Tph:T}
\end{equation}
The physical temperature $\Tph$ depends on $q$ explicitly, 
and $\Tph$ is equal to $T$ when the term including $V$ is negligible.
%%%%%
The quantity $I(T,q)$ is represented as  $I(T,q)=q \Tph^2 / 12 + O((q-1)^2)$
which is obtained  by rewriting the first term of eq.~\eqref{ITq:b} with $\Tph$.
Therefore, the behavior of the physical quantity as a function of $T$ in the case that 
the term including $V$ is negligible is similar to the behavior of the physical quantity as a function of $T_{ph}$.

%%%%%%%%%%%%%%%%%%%%%%%
%%%\subsection{The critical temperature, condensate, and mass}
\section{Numerical estimation}
\label{sec:numericalcalc}   
In this section, we estimate $|\Phi(T,q)/v|$ and $m(T,q)/m(T=0,q)$ numerically.  
We use the ratio $T/v$ and $\Tph/v$ as variables.
We can estimate the critical temperatures for various $q$ from these calculations.

\begin{figure}
\parbox{\figsize}{%
\begin{center}
\includegraphics[bb = 0 0 381 369, width=0.45\textwidth]{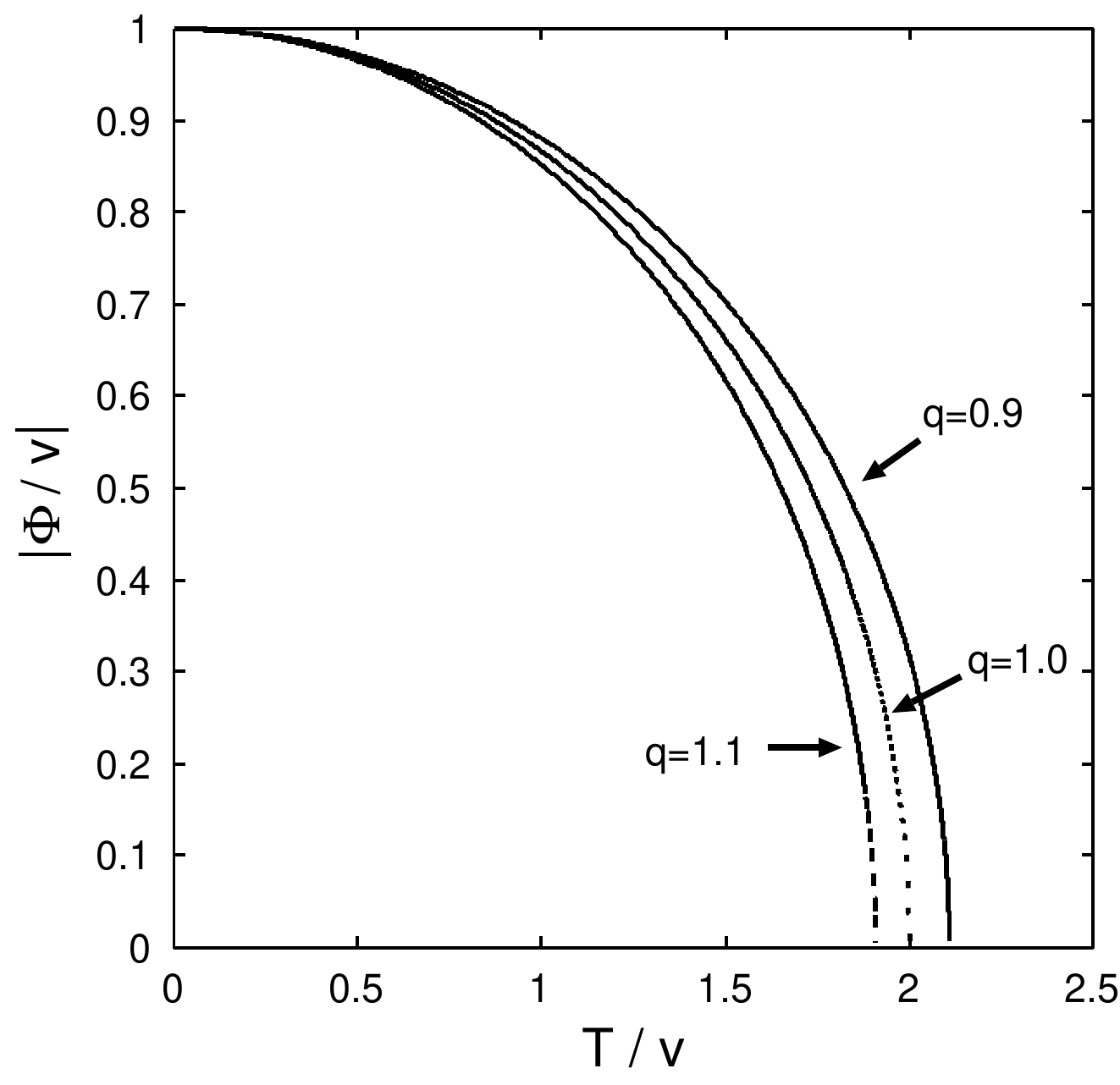}
\end{center}
\caption{The ratio $|\Phi(T,q)/v|$ as a function of $T/v$ for $q=0.9, 1.0,$ and $1.1$, when the term including $V$ in $I(T,q)$ is negligible.}
\label{Fig:Phi-v}
}
%%\end{figure}
%%%%
\hfill
%%%%
%%\begin{figure}
\parbox{\figsize}{%
\begin{center}
\includegraphics[bb = 0 0 389 367, width=0.45\textwidth]{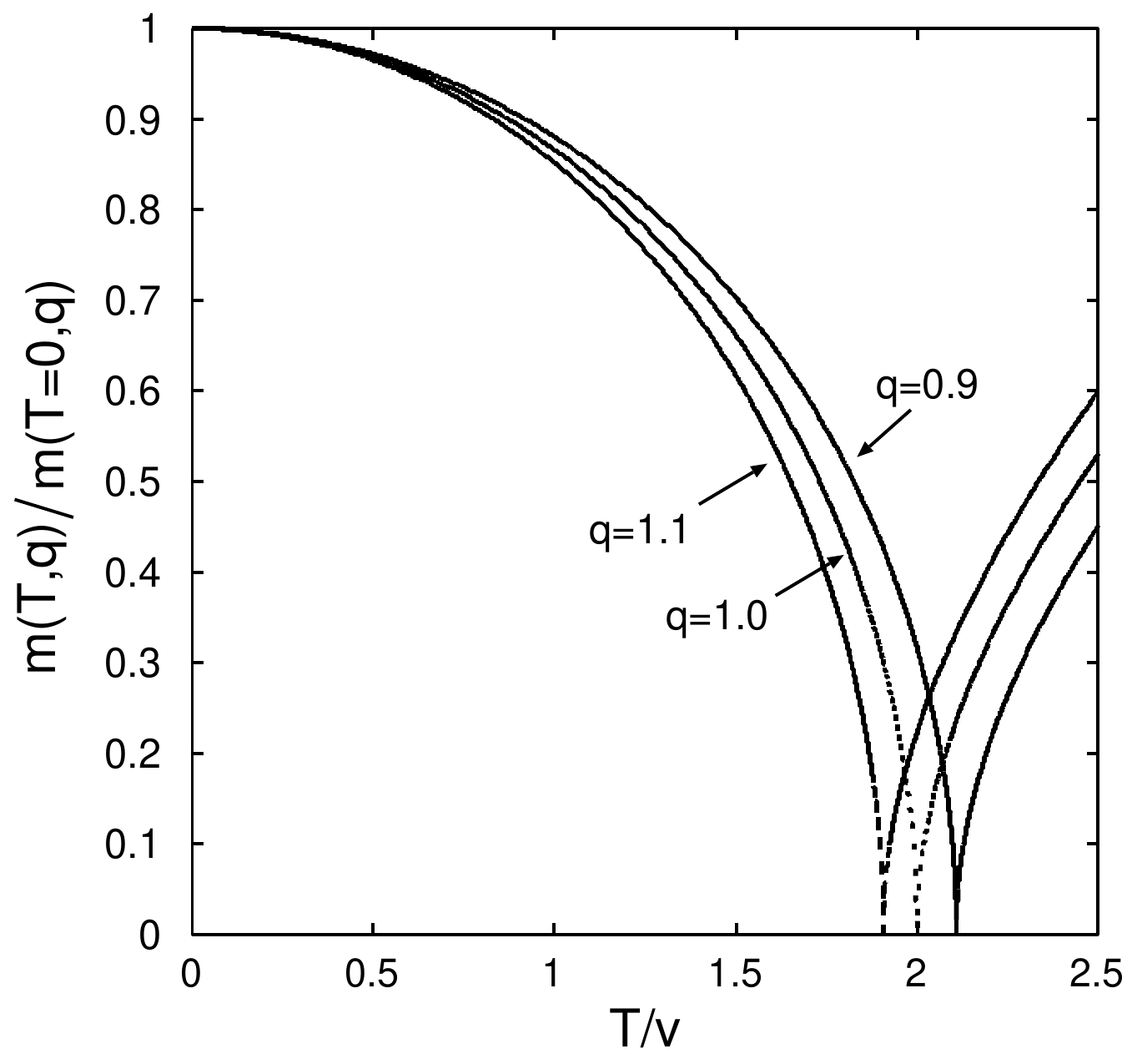}
\end{center}
\caption{The ratio $m(T,q)/m(T=0,q)$ as a function of $T/v$ for $q=0.9, 1.0,$ and $1.1$ when the term including $V$ in $I(T,q)$  is negligible.}
\label{Fig:mass-ratio}
}
\end{figure}

%%%%%%%%
First, we show the numerical results when the term including the volume $V$ 
in the function $I(T,q)$ is negligible (see eq.~\eqref{ITq:b}).
%%%%%%%%%%%%%%%%
Figure~\ref{Fig:Phi-v} shows the quantity $|\Phi(T,q)/v|$ as a function of $T/v$.
The curves are similar in Fig.~\ref{Fig:Phi-v}.
%%%%%%%%%%%%%%%%%%
Figure~\ref{Fig:mass-ratio} shows  the ratio $m(T,q)/m(T=0,q)$ as a function of $T/v$. 
We note that $m(T=0,q)$ is independent of $q$.
As $T$ increases, the mass decreases, reaches minimum, and increases after that. 
The mass at $q>1$  is lighter than the mass at $q=1$ at low temperature,  
and heavier than the mass at $q=1$ at high temperature. 
%%%%% This behavior reflects the behavior of $m(T,q)$ as a function of $T$. 
%%%%%%%%%%%%%%%%
%%%%%%%%
The ratio $T_c(q)/T_c(q=1)$ is simply $1/\sqrt{q}$, and is a monotonically decreasing function of $q$.
The ratio at $q=1.1$ is approximately 0.953 and  the ratio at $q=0.9$ is approximately 1.054. 
This variation is easily understood by expanding $1/\sqrt{1+(q-1)}$ with respect to $q-1$.
This $q$-dependence of $T_c(q)$ is seen in Figs.~\ref{Fig:Phi-v} and \ref{Fig:mass-ratio}.
%%%%%%%%%%%%%%%%%%
%%%%%%%%%%%%%%%%%%

\begin{figure}
\parbox{\figsize}{%
\begin{center}
 \includegraphics[bb = 0 0 384 371, width=0.45\textwidth]{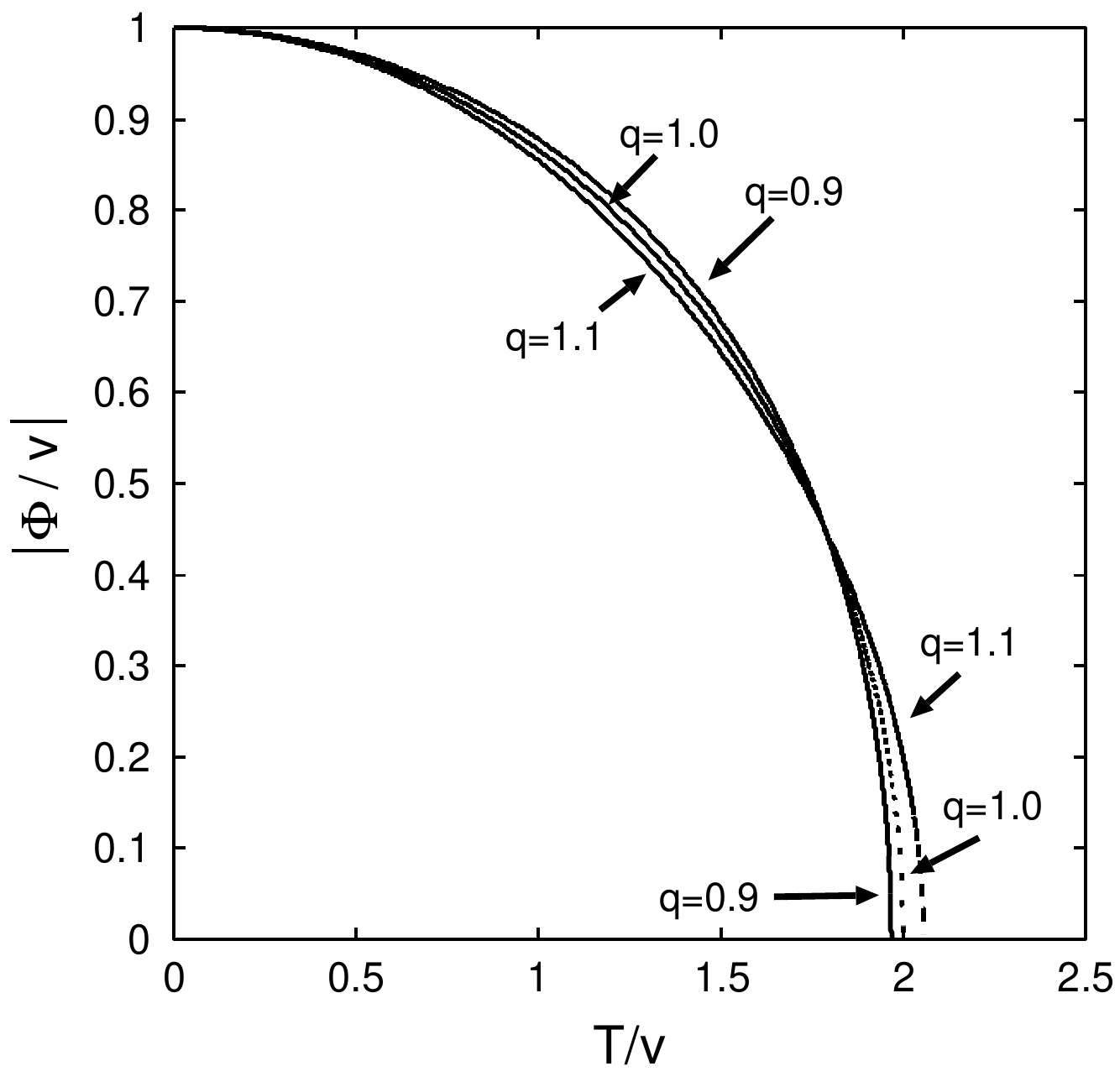}
\end{center}
\caption{The ratio $|\Phi(T,q)/v|$ as a function of $T/v$ at $v^3 V = 0.2$ for $q=0.9, 1.0,$ and $1.1$.}
\label{Fig:Phi-v:finite-V}
}
%%%\end{figure}
\hfill
%%%\begin{figure}
\parbox{\figsize}{%
\begin{center}
\includegraphics[bb = 0 0 387 370, width=0.45\textwidth]{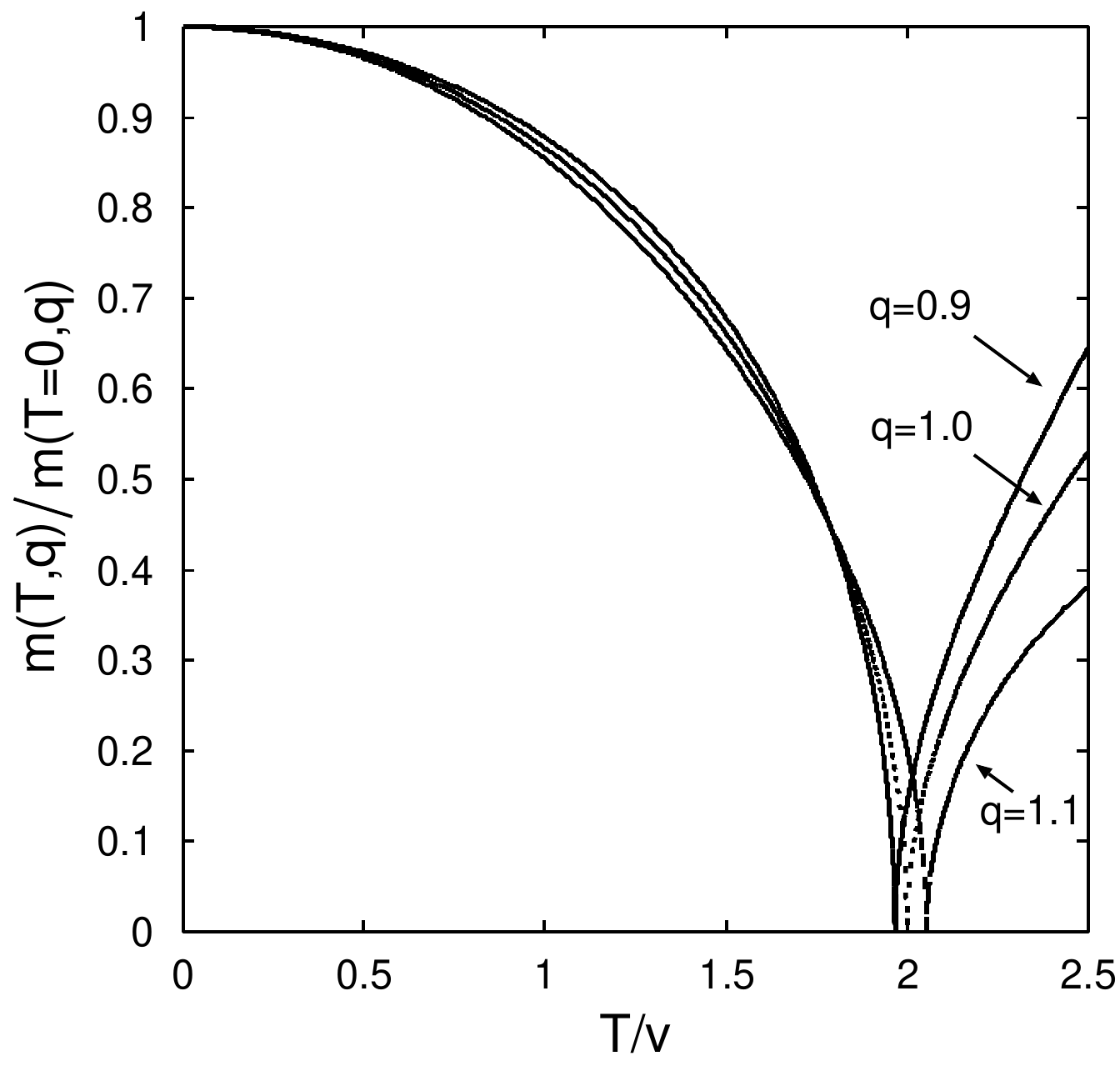}
\end{center}
\caption{The ratio $m(T,q)/m(T=0,q)$ as a function of $T/v$ at $v^3 V = 0.2$ for $q=0.9, 1.0,$ and $1.1$.}
\label{Fig:mass-ratio:finite-V}
}
\end{figure}

%%%%%%%%%%%%%%%%%%%%%%%%%%%%%%%%%%%%
%%%%%%%%%%%%%%%%%%%%%%%%%%%%%%%%%%%%
%%%%%%%%%%%%%%%%%%%%%%%%%%%%%%%%%%%%
Next, we study the quantities numerically with the term including the volume $V$.  %%% I(T,q) の Volume termのつもり
The second term in the square bracket of eq.~\eqref{ITq:a} comes from the coefficient $c_1^M$. 
%%% which is characteristic in the $q$-expectation value.
%%%%%
We set $v^3 V = 0.2$ and calculate some quantities in the range of $0 \le T/v \le 2.5$ for $q=0.9, 1.0$, and $1.1$. 
%%% We note that absolute value of the ratio of the second term of $I(T,q)$ to the first term of $I(T,q)$ is less than 0.25 for $q=1.1$ and $T/v=2.5$.  
%%%%%%%%%%%%%%%%%%
Figure~\ref{Fig:Phi-v:finite-V} shows the quantity $|\Phi(T,q)/v|$ as a function of $T/v$ at $v^3 V = 0.2$ for $q=0.9, 1.0,$ and $1.1$.
The condensate $\Phi(T,q)$ is larger than $\Phi(T,q')$ for $q < q'$ at low temperature.
The $q$-dependence of $\Phi(T,q)/v$ in Fig.~\ref{Fig:Phi-v:finite-V} is similar to that in Fig.~\ref{Fig:Phi-v} at low temperature. 
In contrast, the $q$-dependence of $\Phi(T,q)/v$ in Fig.~\ref{Fig:Phi-v:finite-V} is different from that in Fig.~\ref{Fig:Phi-v} at high temperature. 
The condensate $\Phi(T,q)$ is larger than $\Phi(T,q')$ for $q > q'$ at high temperature.
The critical temperature $T_c(q)$ is larger than $T_c(q')$ for $q > q'$ as shown in Fig.~\ref{Fig:Phi-v:finite-V}, 
while $T_c(q)$ is smaller than $T_c(q')$ for $q > q'$ in Fig.~\ref{Fig:Phi-v}. 
%%%%%%%%%%%%%%%%%%
%%%%%%%%%%%%%%%%%%
Figure~\ref{Fig:mass-ratio:finite-V} shows  the ratio $m(T,q)/m(T=0,q)$ as a function of $T/v$ at $v^3 V = 0.2$ for $q=0.9, 1.0,$ and $1.1$. 
The $q$-dependence of the ratio in Fig.~\ref{Fig:mass-ratio:finite-V} is similar to that in Fig.~\ref{Fig:mass-ratio} at low temperature.
The $q$-dependence of the ratio in Fig.~\ref{Fig:mass-ratio:finite-V} is quite different from that in Fig.~\ref{Fig:mass-ratio} at high temperature.
%%%%%%
%%%%%%
%%%%%%%%%%%%%%%%%%

\begin{figure}
\parbox{\figsize}{%
\begin{center}
\includegraphics[bb = 0 0 384 367, width=0.45\textwidth]{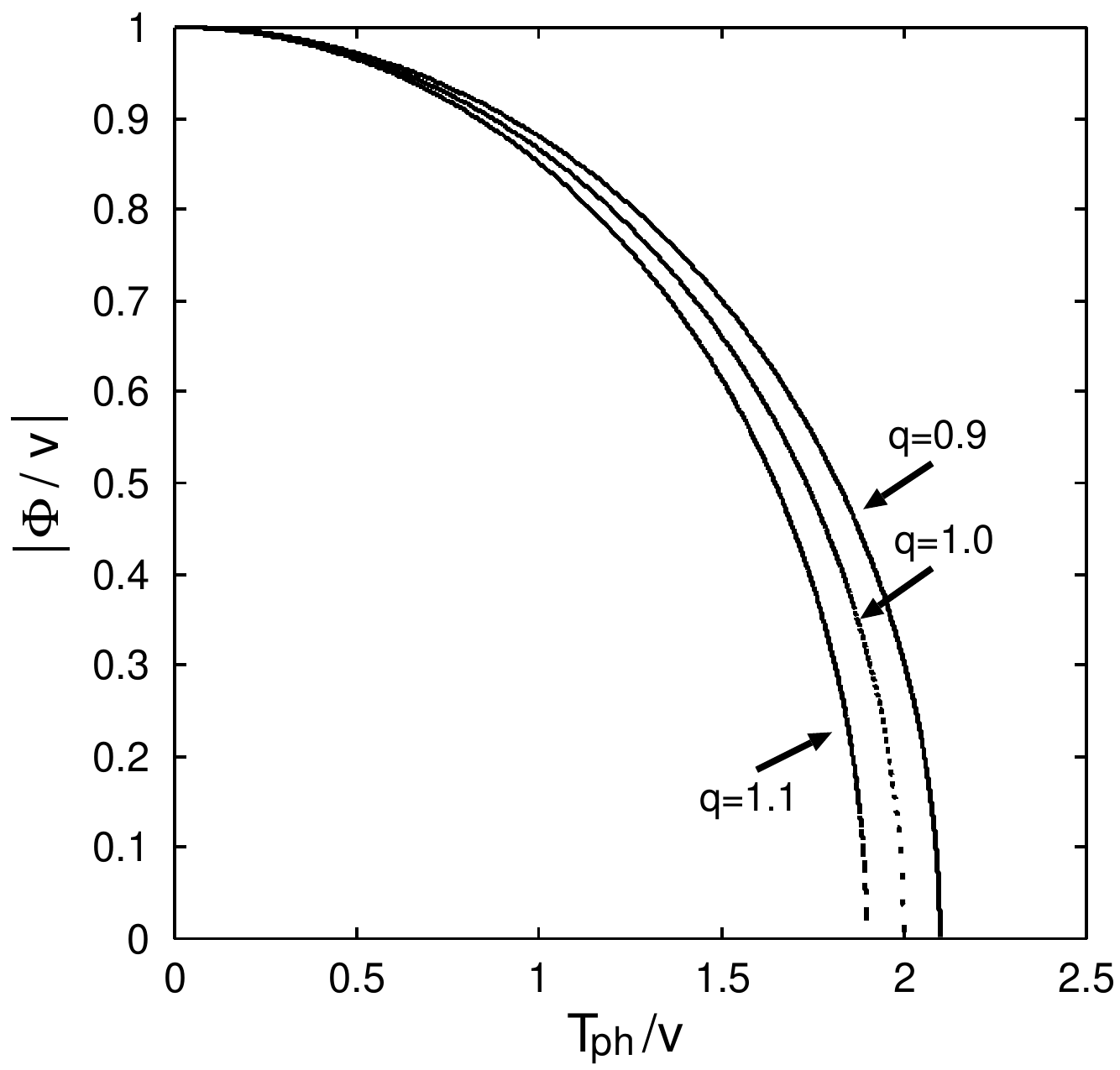}
\end{center}
\caption{The ratio $|\Phi(T,q)/v|$ as a function of $\Tph/v$ at $v^3 V = 0.2$ for $q=0.9, 1.0,$ and $1.1$ in the range of $0 \le T \le 2.5$.}
\label{Fig:Phi-v:finite-V:physical-temperature}
}
%%%\end{figure}
%%%%%%%%%%%%%%%%%%
\hfill
%%%\begin{figure}
\parbox{\figsize}{%
\begin{center}
\includegraphics[bb = 0 0 378 366, width=0.45\textwidth]{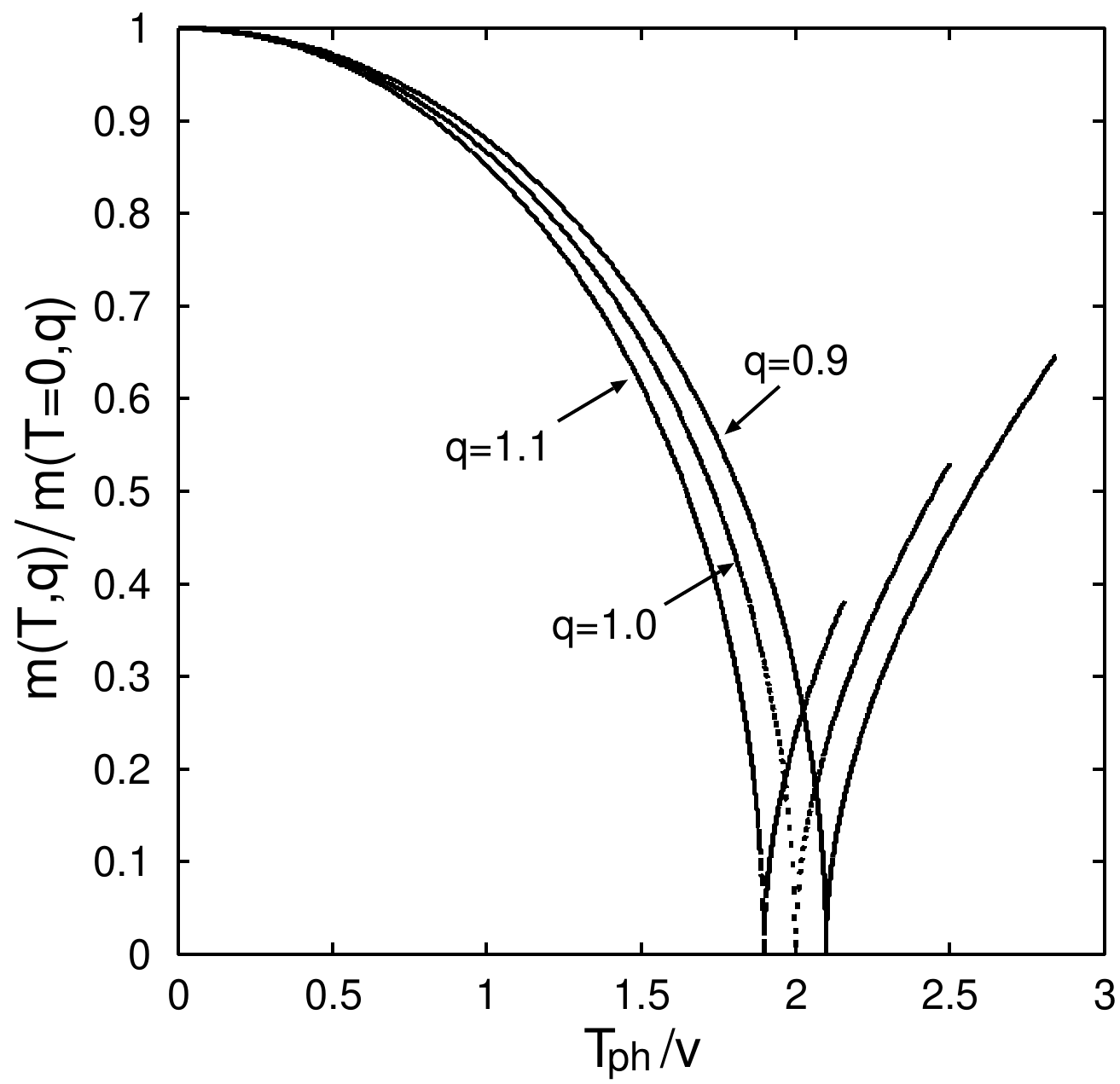}
\end{center}
\caption{The ratio $m(T,q)/m(T=0,q)$ as a function of $\Tph/v$ at $v^3 V = 0.2$ for $q=0.9, 1.0,$ and $1.1$ in the range of $0 \le T \le 2.5$.}
\label{Fig:mass-ratio:finite-V:physical-temperature}
}
\end{figure}

%%%%%%%%%%%%%%%%%%%%%%%
Figure~\ref{Fig:Phi-v:finite-V:physical-temperature} shows $|\Phi(T,q)/v|$ as a function of $\Tph/v$ 
at $v^3 V = 0.2$ for $q=0.9, 1.0,$ and $1.1$ in the range of $0 \le T \le 2.5$.
The points $(\Tph/v, |\Phi/v|)$ were plotted in this figure, because $\Tph$ and $\Phi$ are the functions of $T$.
The behavior of $|\Phi(T,q)/v|$ as a function of $\Tph/v$ in Fig.~\ref{Fig:Phi-v:finite-V:physical-temperature} is similar to that in Fig.~\ref{Fig:Phi-v}.
The critical physical temperature decreases as $q$ increases, %%% at which the condensate becomes zero 
as shown in Fig.~\ref{Fig:Phi-v:finite-V:physical-temperature}.
Figure~\ref{Fig:mass-ratio:finite-V:physical-temperature} shows  the ratio $m(T,q)/m(T=0,q)$ as a function of $\Tph/v$ 
at $v^3 V = 0.2$ for $q=0.9, 1.0,$ and $1.1$ in the range of $0 \le T \le 2.5$. 
In this figure, the points $(\Tph/v, m(T,q)/m(T=0,q))$ were plotted by varying $T$.
The behavior of $m(T,q)/m(T=0,q)$ in Fig.~\ref{Fig:mass-ratio:finite-V:physical-temperature} is also similar to that in Fig.~\ref{Fig:mass-ratio}.
As $\Tph$ increases, the mass decreases, reaches minimum, and increases after that. 
The mass at $q>1$  is lighter than the mass at $q=1$ at low physical temperature,  and heavier than the mass at $q=1$ at high physical temperature. 
%%%%%%%%%%%%%%%%%%%%%%%%
%%%%%%
%%%%%%%%%%%%%%%%%%

%%%%%%%%%%%%%%%%%%%%%%%
\section{Discussion and conclusion}
\label{sec:discussion}
%%%% 
We studied the effects of the nonextensivity on the phase transition for the system of small volume in the $\phi^4$ theory 
in the Tsallis nonextensive statistics of entropic parameter $q$ and temperature $T$.
We adopted the normalized $q$-expectation value.
In this study, the condensate $\Phi(T,q)$ and the mass $m(T,q)$ were calculated for small $|q-1|$ under the massless free particle approximation, 
and the critical temperature was estimated.

%%%%%%%%%%% \beta^{-3} V を無視できる時。
The expressions of these quantities contain the system volume $V$.
The obtained $q$-dependences of the quantities without the terms including $V$ are probable 
when the term including $V$ in $\qexpect{\norder{\varphi^2}}$ is negligible,  
because it is expected that $\qexpect{\norder{\varphi^2}}$ at $q>1$ is larger than that at $q=1$. 
Indeed,  the critical temperature is a monotonically decreasing function of $q$
when the term including $V$ is negligible, as shown in eq.~\eqref{eqn:Tc}.
The critical temperature, the condensate, and the mass are the functions of the effective temperature $T^* = \sqrt{q} T$.

%%%%%%%%%%% \beta^{-3} V を無視できなくなると・・・
The $q$-dependences of the quantities show different behaviors when the terms including $V$ are not negligible. 
The corrections work at large $T/v$, as shown in Fig.~\ref{Fig:Phi-v:finite-V} and Fig.~\ref{Fig:mass-ratio:finite-V}.
In particular, the $q$-dependence of the quantity with the term including $V$ differs from that without the term including $V$, 
when term including $V$ is sufficiently large. 
The term including $V$ comes from the coefficient $c_1^M$.
This fact indicates that 
the definition of the expectation value and the constraints for the density operator are significant 
for the $q$-dependence of a physical quantity
when the volume of the system is not sufficiently small.

%%%%%%%%%%% Phyical temperatureを用いると %%%%%%%%%%%
The behavior of $\Phi/v$ with the term including $V$ as a function of $\Tph/v$ (Fig.~\ref{Fig:Phi-v:finite-V:physical-temperature})
is similar to that without the term including $V$ as a function of $T/v$ (Fig.~\ref{Fig:Phi-v}). 
The behavior of $m(T, q)/m(T=0, q)$  with the term including $V$ as a function of $\Tph/v$ (Fig.~\ref{Fig:mass-ratio:finite-V:physical-temperature})
is also similar to that without the term including $V$ as a function of $T/v$ (Fig.~\ref{Fig:mass-ratio}).
The quantity $\Phi/v$ at $q$ is smaller than that at $q'$ for $q>q'$.
The $q$-dependence of $\Phi/v$ is valid, 
because it is expected that the contribution of the distribution at $q$ is  larger than that at $q'$ for $q>q'$.
The $q$-dependences of the mass is also valid.
The similarity is explained by the expression $I(T,q)=q \Tph^2 / 12 + O((q-1)^2)$. 
These behaviors indicate probably the significance of the physical temperature, 
and imply probably the importance of the definition of the expectation value and the constraints for the density operator.  

%%%% Summary of conclusion
In summary, 
we studied the effects of the nonextensivity on the phase transition for the system of small volume $V$ in the $\phi^4$ theory 
in the Tsallis nonextensive statistics of small $|q-1|$, where the quantity $q$ is the entropic parameter.
We adopted the normalized $q$-expectation value.
We calculated the condensate and the mass to the order $q-1$ under the massless free particle approximation.
%%% 
The condensate $\Phi(T,q)$ divided by $v$, $\Phi/v$, at $q$ is smaller than that at $q'$ for $q>q'$ 
as a function of $\Tph/v$ which is the physical temperature $\Tph$ divided by $v$. 
The mass decreases, reaches minimum, and increases after that, as $\Tph$ increases. 
The mass at $q>1$  is lighter than the mass at $q=1$ at low physical temperature,  and heavier than the mass at $q=1$ at high physical temperature,
as a function of $\Tph$. 
%%%
The effects of the nonextensivity on the physical quantity as a function of $\Tph$ become strong as the quantity $|q-1|$ increases. 
As functions of $T$, the $q$-dependence of the quantity with the term including $V$ differs from that without the term including $V$.
The difference is large at high temperature $T$.
The $q$-dependence of the quantity with the term including $V$ as a function of $\Tph$ 
is similar to that without the term including $V$ as a function of $T$. 
%%%
These $q$-dependences indicate that 
the definition of the physical temperature, the definition of the expectation value, and the constraints for the density operator are significant
in the Tsallis nonextensive statistics when the volume of the system is not sufficiently small.
%%%%% even though the volume is small. 

%%%% outlook
We hope that this work is helpful to understand the effects of the nonextensivity 
with the normalized $q$-expectation value in field theories.

%%%%%%%%%%%%%%%%%%%%%%%
\appendix
\section{Traces}
\label{app:sec:trace}
The following traces appear in the calculations: 
%%%%%%%%%%%%%%%
\begin{subequations}
\begin{align}
& \Lambda_{n} := \trace{ \exp(-\beta \nHM) (\nHM)^n } ,
\\
& \Lambda_{n,\vec{k}} := \trace{ \exp(-\beta \nHM ) (\nHM)^n a^{\dag}_{\vec{k}} a_{\vec{k}}} ,
\end{align}
\label{Lambda}
\end{subequations}
%%%%%%%%%%%%%%%
where $a_{\vec{k}}$ is the annihilation operator, and $:H^M:$ is the normal ordered Hamiltonian of a free field. 
That is, 
\begin{align}
\nHM = \sum_{\vec{l}}  \omega_{\vec{l}}  a^{\dag}_{\vec{l}}  a_{\vec{l}},
\end{align}
where $\omega_{\vec{l}}$ is the energy of a particle with momentum $\vec{l}$. 
It is easily found from the definitions, eq.~\eqref{Lambda}, that 
$\Lambda_{n}$ and $\Lambda_{n,\vec{k}}$ satisfy the following relation:
\begin{align}
\Lambda_{n+1} = \sum_{\vec{k}} \omega_{\vec{k}} \Lambda_{n,\vec{k}} .
\end{align}

We give the expressions of $\Lambda_n$ and $\Lambda_{n, \vec{k}}$ explicitly.
\begin{subequations}
\begin{align}
& \Lambda_0 =  \prod_{\vec{l}} \left( 1 - e^{- \beta \omega_{\vec{l}}} \right)^{-1}  ,\\
& \Lambda_1 =  \left[ \sum_{\vec{l}} \frac{\omega_{\vec{l}}}{\left( e^{\beta \omega_{\vec{l}}} -1 \right)} \right] \Lambda_0   ,\\
& \Lambda_2 =  \left\{ 
\left[ \sum_{\vec{l}} \frac{\left( \omega_{\vec{l}} \right)^2  e^{\beta \omega_{\vec{l}}} }{\left( e^{\beta \omega_{\vec{l}}} -1 \right)^2} \right]
+
\left[ \sum_{\vec{l}} \frac{\omega_{\vec{l}}}{\left( e^{\beta \omega_{\vec{l}}}  -1 \right) } \right]^2 
\right\} \Lambda_0  ,\\
%%%
&  \Lambda_{3}  =  \left\{ 
\left[ \sum_{\vec{l}} \frac{\left( \omega_{\vec{l}} \right)^3 e^{\beta \omega_{\vec{l}}}\left( e^{\beta \omega_{\vec{l}}} + 1 \right) }{\left( e^{\beta \omega_{\vec{l}}} - 1 \right)^3} \right]
+
3 \left[ \sum_{\vec{l}} \frac{\omega_{\vec{l}}}{\left( e^{\beta \omega_{\vec{l}}} -1 \right)} \right]
\left[ \sum_{\vec{l}} \frac{\left( \omega_{\vec{l}}\right)^2 e^{\beta \omega_{\vec{l}}} }{\left( e^{\beta \omega_{\vec{l}}} - 1 \right)^2}  \right] 
+ 
 \left[ \sum_{\vec{l}} \frac{\omega_{\vec{l}}}{\left( e^{\beta \omega_{\vec{l}}} -1 \right)} \right]^3
\right\}
\Lambda_{0}
, \\
%%%%%%%%%%%
& \Lambda_{0,\vec{k}}  =   \frac{1}{\left( e^{\beta \omega_{\vec{k}}} - 1 \right)} \Lambda_0  ,\\
& \Lambda_{1,\vec{k}}  =   \left\{ 
\frac{\omega_{\vec{k}} e^{\beta \omega_{\vec{k}}}}{\left( e^{\beta \omega_{\vec{k}}} - 1 \right)} 
+ 
 \left[ \sum_{\vec{l}} \frac{\omega_{\vec{l}}}{\left( e^{\beta \omega_{\vec{l}}} -1 \right)} \right]
\right\}
\Lambda_{0,\vec{k}}  ,\\
& \Lambda_{2,\vec{k}}  =   \left\{ 
\frac{\left( \omega_{\vec{k}} \right)^2 e^{\beta \omega_{\vec{k}}}\left( e^{\beta \omega_{\vec{k}}} + 1 \right) }{\left( e^{\beta \omega_{\vec{k}}} - 1 \right)^2}
+
\left[ \frac{2 \omega_{\vec{k}} e^{\beta \omega_{\vec{k}}} }{\left( e^{\beta \omega_{\vec{k}}} - 1 \right)}  \right] 
\left[ \sum_{\vec{l}} \frac{\omega_{\vec{l}}}{\left( e^{\beta \omega_{\vec{l}}} -1 \right)} \right]
\right. \nonumber \\  & \qquad\qquad \left. 
+ 
 \left[ \sum_{\vec{l}} \frac{\omega_{\vec{l}}}{\left( e^{\beta \omega_{\vec{l}}} -1 \right)} \right]^2
+ 
 \left[ \sum_{\vec{l}} \frac{\left( \omega_{\vec{l}} \right)^2 e^{\beta \omega_{\vec{l}}} }{\left( e^{\beta \omega_{\vec{l}}} -1 \right)^2} \right]
\right\}
\Lambda_{0,\vec{k}}
.
\end{align}
\end{subequations}

%%%%%%%%%%%%%%%%%%%%%%%%%%%%%%%%%%%%%
\section{Integrals}
\label{app:sec:integrals}
Some integrals appear in the calculations. 
In this appendix, we show the results of the integrals. 
The following integral appears:
\begin{align}
\int_0^{\infty} \ dx \ x^m \ln(1-e^{-x}) = - \Gamma(m+1) \zeta(m+2) \qquad (m\ge 0, m \in \mathbb{N}) , 
\end{align}
where $\zeta$ is the zeta function.

%%%%%
Another integral appears:
%%%%%%%%%%%%%%%
\begin{subequations}
\begin{align}
F(a,b,c; \nu) := \int_0^{\infty} \ dx \ x^c \frac{\left( e^{x} \right)^b}{\left( e^{x+\nu} - 1 \right)^a }  \quad (a, b, c \in \mathbb{N})
%%% \qquad (a>  b \ge 0, c \ge 0 ;  \quad a, b, c \in \mathbb{N}, \nu \in \mathbb{R} )
.
\end{align}
\end{subequations}
%%%%%%%%%%%%%%%
This integral is represented as follows (when the sum converges):  
\begin{align}
F(a,b,c; \nu) 
= e^{-b \nu} \Gamma(c+1) \sum_{N_a=0}^{\infty} \frac{e^{- (N_a + a - b) \nu}}{(N_a + a - b)^{c+1}} 
\sum_{N_{a-1}=0}^{N_a}  \sum_{N_{a-2}=0}^{N_{a-1}} \cdots \sum_{N_{1}=0}^{N_2} 1 
.
\end{align}
We now focus on the integrals for $\nu=0$: %%% the case of $\nu=0$:
\begin{align}
F(a,b,c; \nu=0) 
= \Gamma(c+1) \sum_{N_a=0}^{\infty} \frac{1}{(N_a + a - b)^{c+1}} \sum_{N_{a-1}=0}^{N_a}  \sum_{N_{a-2}=0}^{N_{a-1}} \cdots \sum_{N_{1}=0}^{N_2} 1 
. 
\end{align}
The explicit results of some integrals are shown below.
%%%%%%%%%%%%%%%
\begin{subequations}
\begin{align}
& F(1, 0 , c; \nu=0) = \Gamma(c+1) \zeta(c+1) , \\
& F(2, 0 , c; \nu=0) = \Gamma(c+1) \left[ \zeta(c)  - \zeta(c+1) \right] , \\
& F(2, 1 , c; \nu=0) =  \Gamma(c+1) \zeta(c) , \\
& F(3, 0 , c; \nu=0) = \frac{1}{2} \ \Gamma(c+1) \left[ \zeta(c-1)  - 3 \zeta(c) + 2 \zeta(c+1) \right] , \\
& F(3, 1 , c; \nu=0) = \frac{1}{2} \ \Gamma(c+1) \left[ \zeta(c-1)  - \zeta(c) \right] , \\
& F(3, 2 , c; \nu=0) = \frac{1}{2} \ \Gamma(c+1) \left[ \zeta(c-1)  + \zeta(c) \right] .
\end{align}
\end{subequations}
%%%%%%%%%%%%%%%

The function $F(a, b, c; \nu)$ is represented with the appell function $\phi(z, s)$ \cite{Book:Iwanami3} for $\nu >0$.
For example, the functions $F(1, 0 , c; \nu)$ and $F(2, 1 , c; \nu)$ are given by 
%%%%%%
\begin{subequations}
\begin{align}
& F(1, 0 , c; \nu) = \Gamma(c+1) \phi(c+1, e^{-\nu})  , \\
& F(2, 1 , c; \nu) = e^{-\nu} \Gamma(c+1) \phi(c, e^{-\nu}) . 
\end{align}
\end{subequations}
These equations may be useful in the future studies.  

%%%%%%%%%%%%%%%%%%

%%%%%%%%%%%%%%%%%%%%%%%

\end{document}